\RequirePackage{fix-cm}
\documentclass[twocolumn,smallextended]{svjour3}       % onecolumn (second format)
\usepackage{color}
\smartqed  % flush right qed marks, e.g. at end of proof
\usepackage{graphicx}
\journalname{Brazilian Journal of Physics}
\begin{document}

\title{ Transformation process of the magnetron-sputtered Ag$_2$O film in hydrogen annealing%\thanks{Grants or other notes
}


\author{Xiaoyong GAO \and Mengke ZHAO \and Hongtao LIU \and Sa ZHANG
         %etc.
}


\institute{Xiaoyong GAO  \at
              Key Laboratory of Material Physics of Ministry of Education, School of Physics and Engineering, Zhengzhou University, Zhengzhou 450052, China \\
              Tel.: +086-371-67766917\\
              Fax: +086-371-67767803\\
              \email{xygao@zzu.edu.cn}           %  \\
           %\and
           %S. Author \at
              %second address
}

\date{Received: date / Accepted: date}

\maketitle

\begin{abstract}
The current paper mainly addresses the effect of the hydrogen partial pressure on the microstructure and transformation of the Ag$_2$O film. The transformation process and mechanism were also analyzed in detail. Increasing the hydrogen partial pressure can accelerate the transformation of Ag$_2$O to Ag and lower the critical transformation temperature of the film due to the enhanced hydrogen reduction, and to both of the lowered activation energy of the reaction of Ag$_2$O with hydrogen and enhanced lattice strain of the Ag$_2$O film, respectively. Hydrogen-involved reaction in the whole hydrogen annealing process is mainly hydrogen reduction reaction with Ag$_2$O. The diffusion of hydrogen and gaseous H$_2$O molecules is accompanied with the whole hydrogen annealing process.
\end{abstract}

\section{Introduction}
\label{intro}
  Ag$_x$O films have attracted considerable attention because of their potential application in optical and magneto-optical storage. In 1992, Tominaga et al. \cite{1} first reported that Ag$_x$O films can be used as substitutes for the organic material commonly used as the storage material in recordable compact discs (CD-R). Chiu \cite{2} and Kim et al. \cite{3} suggested that Ag$_x$O can be used as a mask layer in a magneto-optical disc to produce a high-resolution aperture and enhance the magneto-optical signals. In 2003, Zhang et al. \cite{4} discovered the dynamically activated luminescence emission of Ag$_x$O films, proving that Ag$_x$O films can be used as nanoscale optical storage materials. According to previous reports, Ag$_x$O contains several phases, namely, AgO, Ag$_2$O, Ag$_3$O$_4$ \cite{5}, Ag$_4$O$_3$ \cite{6}, and Ag$_2$O$_3$, of which Ag$_2$O is the most thermodynamically stable. The magnetron-sputtered Ag$_x$O film is usually biphased (including AgO and Ag$_2$O phases). However, the use of the Ag$_x$O film as a new-generation disc material depends on whether it can be used as an optical switch. In 2006, an experiment was performed by Qin et al. \cite{7} to study the ablation of the Ag$_x$O film under different laser powers. The results suggested that the Ag$_x$O film can be used as an optical and magneto-optical storage material mainly because of its thermal decomposition reaction. Chuang et al. \cite{8} pointed out that the critical thermal decomposition temperature (CDT) of the AgO and Ag$_2$O phases are 160 and 380 $^o$C, respectively, whereas Abe et al. \cite{9} found that the CDT of the Ag$_2$O film deposited via radio-frequency magnetron reactive sputtering is between 200 and 400 $^o$C. Gao et al. \cite{10} previously reported that the Ag$_x$O film deposited via direct-current magnetron reactive sputtering (DC sputtering) consists of AgO and Ag$_2$O phases, with CDTs at 200 and 300 $^o$C, respectively. As recently reported by Gao et al. \cite{11}, a $<$111$>$ preferentially oriented Ag$_2$O film is deposited on a glass substrate via DC sputtering. The Ag$_2$O film was then annealed using an optical excitation-assisted rapid thermal annealing under a nitrogen protection condition \cite{12} and a traditional chamber annealing under a vacuum condition \cite{13} to determine its thermal stability. In the annealing process, Ag$_2$O begins to transform into Ag and O$_2$ at a CDT of approximately 200 $^o$C. In addition, O$_2$ diffusion is accompanied by the thermal decomposition of Ag$_2$O.

Hydrogen annealing may play different roles in different films. Liu et al. \cite{14} reported that the electrical property and crystal quality of ZnO films significantly changed after hydrogen annealing. Han et al. \cite{15} reported that hydrogen atoms interstitially couple to oxygen and form hydroxyl (HO)$^-$ ions, resulting in the loss of oxygen during the hydrogen annealing of the Ti$_{0.93}$O$_3$ film. Gao et al. \cite{16} reported that the hydrogen reduction effect can lower the critical transformation temperature (CTT) of Ag$_2$O to 175 $^o$C at a hydrogen partial pressure (HPP = [H$_2$ ]/([Ar]+ [H$_2$])) of 25\%. However, no systematic investigation on the effect of HPP on the microstructure and transformation of the hydrogen-annealed Ag$_2$O films has been conducted. In addition, using hydrogen annealing is effective to obtain the information on the transformation process and mechanism of Ag$_2$O. Hence, the current paper focuses on the effect of HPP on the microstructure and transformation of the Ag$_2$O film and the roles that hydrogen plays in hydrogen annealing to obtain deeper insight into the transformation process and mechanism of the Ag$_2$O film.

\section{Experiment}
\label{sec:1}
%Text with citations \cite{RefB} and \cite{RefJ}.
A single-phase, $<$111$>$-oriented Ag$_2$O film was first deposited on glass substrates via DC magnetron sputtering at a working pressure of 2.5 Pa, a substrate temperature of 250 $^o$C, and an O$_2$-to-Ar flow ratio of 15:18. A high-purity (99.995\%) silver plate was used as the sputtering target. During film deposition, the sputtering power was maintained at 95 W and the electrode distance at 35 mm. Prior to film preparation, the glass substrate was rinsed sequentially with acetone and alcohol in an ultrasonic bath for 15 minutes, and then placed in an ultra-high vacuum chamber previously evacuated to a base pressure below 1$\times$10$^{-4}$ Pa. The thickness of the as-deposited Ag$_2$O film is approximately 920 nm. The Ag target was first pre-sputtered for 5 min before the preparation of Ag$_2$O film. The gas flow ratio of oxygen to argon was accurately controlled by a mass flow controller. The injected oxygen and Ar were first mixed in a mixer and then injected to the chamber. In addition, the substrate temperature and working pressure were controlled using a thermal coupler and a high-precision vacuum meter,respectively. The as-deposited Ag$_2$O film was then annealed for one hour at different HPPs using different hydrogen annealing temperatures (T$_a$). The chosen HPP values in the hydrogen annealing are 10\%, 20\%, 30\%, 40\%, and 50\%. The working pressure during hydrogen annealing was maintained at 133 Pa. Afterwards, the films were naturally cooled to room temperature.

The crystalline structure of the film was measured using an X-ray diffractometer (Philips PANalytical X'pert) with a CuK$_{\alpha}$ ($\lambda$ = 0.1540598 nm) as the radiation source. The surface morphology was determined using a cold field scanning electron microscope (JSM-6700F). All measurements were conducted at room temperature.

\section{Results and discussion}
\begin{figure}
  \includegraphics[width=0.45\textwidth]{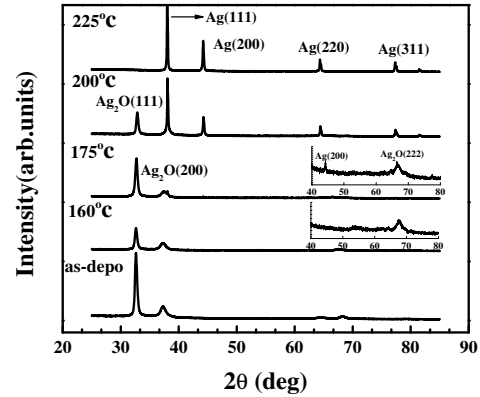}
\caption{X-ray diffraction patterns of the Ag$_2$O film hydrogen annealed at different T$_a$ values using an HPP of 10\%. The two insets denote the X-ray diffraction patterns in the 2$\theta$ ranging from 40$^o$ to 80$^o$.}
\label{fig:1}       % Give a unique label
\end{figure}
Figure~\ref{fig:1} shows the X-ray diffraction patterns of the Ag$_2$O film hydrogen annealed at different T$_a$ values and at an HPP of 10\%. Wei et al.\cite{17}pointed out that the electrodeposited silver oxide nanostructures have a defective cubic structure containing also amorphous zones. However, no defective cubic structure containing any amorphous zones for the magnetron-sputtered Ag$_2$O film is observed. The Ag$_2$O (111) diffraction peak is obviously weakened at T$_a$ = 160 $^o$C, and no Ag diffraction peak is discernible. Two weak Ag (111) and (200) diffraction peaks begin to appear at T$_a$ = 175 $^o$C, indicating that Ag$_2$O begins to transform into Ag at this temperature at an HPP of 10\%. Zhang et al. \cite{13} previously reported that under a vacuum condition, the Ag$_2$O film begins to thermally decompose into Ag nanoparticles at T$_a$ = 300 $^o$C via chamber annealing, indicating that hydrogen annealing can lower the CTT of the Ag$_2$O film. Ag diffraction peaks, rather than Ag$_2$O diffraction peaks, are discerned at T$_a$ = 225 $^o$C.
\begin{figure}
  \includegraphics[width=0.45\textwidth]{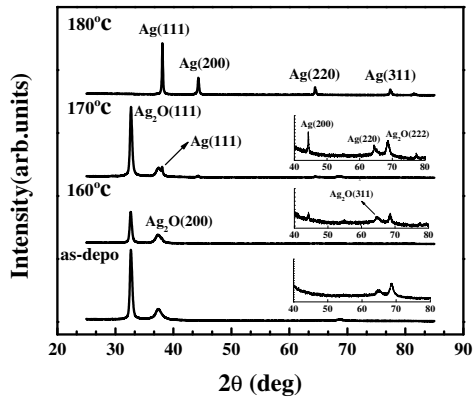}
\caption{X-ray diffraction patterns of the Ag$_2$O film hydrogen annealed at different T$_a$ values using an HPP of 20\%. The three insets denote the X-ray diffraction patterns in the 2$\theta$ ranging from 40$^o$ to 80$^o$.}
\label{fig:2}       % Give a unique label
\end{figure}
\begin{figure}
  \includegraphics[width=0.45\textwidth]{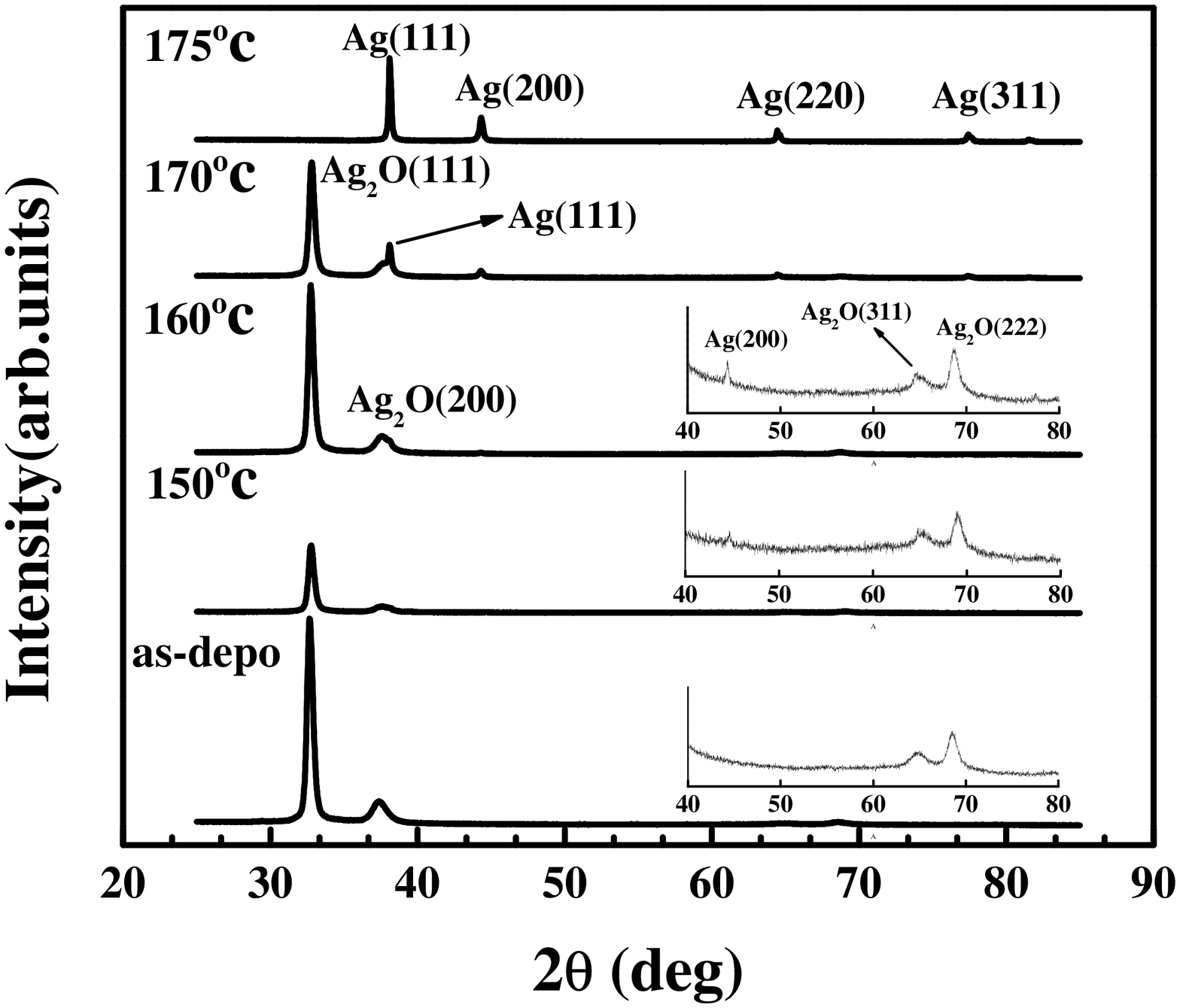}
\caption{X-ray diffraction patterns of the Ag$_2$O film hydrogen annealed at different T$_a$ values using an HPP of 30\%. The three insets denote the X-ray diffraction patterns in the 2$\theta$ ranging from 40$^o$ to 80$^o$.}
\label{fig:3}       % Give a unique label
\end{figure}
\begin{figure}
  \includegraphics[width=0.45\textwidth]{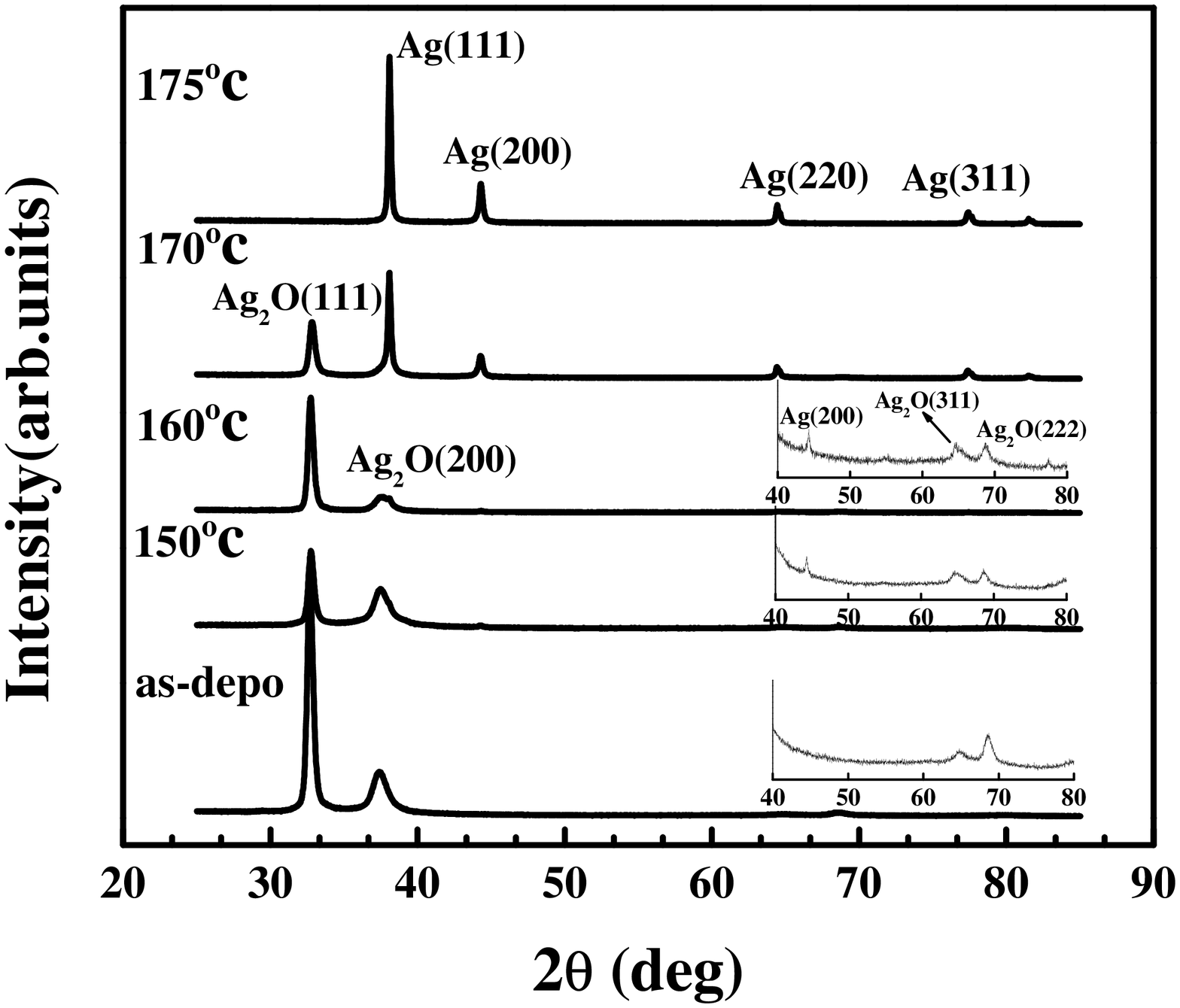}
\caption{X-ray diffraction patterns of the Ag$_2$O film hydrogen annealed at different T$_a$ values using an HPP of 40\%. The three insets denote the X-ray diffraction patterns in the 2$\theta$ ranging from 40$^o$ to 80$^o$.}
\label{fig:4}       % Give a unique label
\end{figure}
\begin{figure}
  \includegraphics[width=0.45\textwidth]{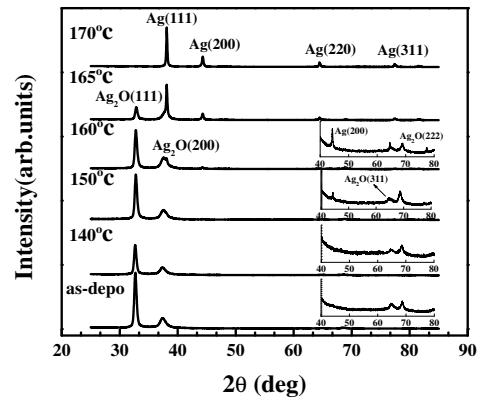}
\caption{X-ray diffraction patterns of the Ag$_2$O film hydrogen annealed at different T$_a$ values using an HPP of 50\%. The four insets denote the X-ray diffraction patterns in the 2$\theta$ ranging from 40$^o$ to 80$^o$.}
\label{fig:5}       % Give a unique label
\end{figure}

Figure~\ref{fig:2} shows the X-ray diffraction patterns of the Ag$_2$O film hydrogen annealed at different T$_a$ values using an HPP of 20\%. A similar evolution from Ag$_2$O to a biphased Ag$_2$O and Ag complex and to Ag occurs as T$_a$ increases. However, a weak Ag (200) diffraction peak begins to appear at T$_a$ = 160 $^o$C. Meanwhile, only Ag diffraction peaks are discerned when T$_a$ is increased to 180 $^o$C. Figure~\ref{fig:3} shows the X-ray diffraction patterns of the hydrogen annealed Ag$_2$O film using an HPP of 30\%. A very weak but clear Ag (200) diffraction peak begins to appear at T$_a$ = 150 $^o$C and becomes strong at T$_a$ = 160 $^o$C, at which the Ag$_2$O diffraction peak is obviously weakened. The Ag$_2$O diffraction peaks completely disappear at T$_a$ = 175 $^o$C. According to the X-ray diffraction results in Figs. ~\ref{fig:1} to ~\ref{fig:3}, the CTT of the Ag$_2$O film is reduced from 175 $^o$C to 150 $^o$C as the HPP increases from 10\% to 30\%, indicating that increasing the HPP can lower down the CTT of the Ag$_2$O film. The lowered CTT value with increasing HPP is inherently attributed to the lowered activation energy of the reaction of Ag$_2$O with hydrogen \cite{18}. In fact, this is a consequence of a great drop in free energy of the reaction system caused by the formation of water \cite{18}. To confirm this result, the Ag$_2$O film was hydrogen annealed using an HPP of 40\% and 50\%. The X-ray diffraction patterns of the Ag$_2$O films are shown in Figs.~\ref{fig:4} and~\ref{fig:5}, respectively. A weak Ag (200) diffraction peak begins to appears at T$_a$ = 150 $^o$C (Fig. ~\ref{fig:4}), accompanied by a weakened Ag$_2$O diffraction peak. The Ag diffraction peaks become much stronger than the Ag$_2$O diffraction peaks at T$_a$ = 170 $^o$C, and only Ag diffraction peaks are discernible at T$_a$ = 175 $^o$C. Figure~\ref{fig:5} shows the appearance of the Ag (200) diffraction peak at T$_a$ = 150 $^o$C, whereas no Ag diffraction peak appears at T$_a$ = 140 $^o$C. Notably, only Ag diffraction peaks are discernible at T$_a$ = 170 $^o$C. The X-ray diffraction results in Figs.~\ref{fig:4} and ~\ref{fig:5} indicate that the CTT of the Ag$_2$O film is maintained at 150 $^o$C as the HPP exceeds 30\%. Compared with the results in Figs.~\ref{fig:1} to ~\ref{fig:3}, increasing the HPP can accelerate the transformation of Ag$_2$O to Ag, which is attributed to the reduction role that hydrogen plays in the hydrogen annealing of the Ag$_2$O film.

Two effects may  coexsit during the hydrogen annealing process. One is the thermal decomposition of the film because of the thermal effect, and the other one is film reduction due to the hydrogen reduction effect. However, the former occurs only at T$_a$ $\ge$ 200 $^o$C \cite{16} in the absence of hydrogen. The former may not occur in parallel to the reduction reaction due to the slow thermal decomposition. On the other hand, the enthalpy of adsorption of H atoms on Ag$_2$O surface exceeds the enthalpy of dissociation of H$_2$ molecule. Adsorption of H$_2$ molecules can not provide enough energy to proceed the dissociation of H$_2$ molecule on Ag$_2$O surface\cite{19}. Thereafter,the hydrogen reduction effect involved in the entire hydrogen annealing process is described as follows:

%\subsection{Subsection title}
%\label{sec:2}
%as required. Don't forget to give each section
%and subsection a unique label (see Sect.~\ref{sec:1}).
%\paragraph{Paragraph headings} Use paragraph headings as needed.

\begin{equation}
Ag_2O +  H_2 (g) = 2 Ag (s) + H_2O (g)
\label{eq:1}
\end{equation}

 When HPP is increased, more H$_2$ molecules may be adsorbed on the Ag$_2$O film surface to accelerate the transformation of Ag$_2$O to Ag, causing the reduction in the CTT. However, the concentration of the H$_2$ molecules adsorbed on the film surface may increase to saturation when HPP exceeds 30\%. The transformation of Ag$_2$O to Ag will no longer be accelerated, causing the CTT to remain unchanged even when HPP further increases.
\begin{table}
\caption{Calculated average grain size and lattice strain of the Ag$_2$O film.}
\label{tab:1}       % Give a unique label
\begin{tabular}{llll}
\hline\noalign{\smallskip}
HPP (\%) & T$_a$($^o$C) & Grain size (nm) & Lattice strain (\%)  \\
\noalign{\smallskip}\hline\noalign{\smallskip}
20 & as-depo & 61.8 & 0.315 \\
20 & 160 & 54.8 & 0.344 \\
20 & 170 & 45.0 & 0.395 \\
30 & as-depo & 38.0 & 0.450 \\
30 & 150 & 35.2 & 0.476 \\
30 & 160 & 32.9 & 0.500 \\
40 & as-depo & 38.0 & 0.448 \\
40 & 150 & 38.0 & 0.449 \\
40 & 160 & 30.9 & 0.526 \\
50 & as-depo & 54.8 & 0.343 \\
50 & 140 & 45.0 & 0.394 \\
50 & 150 & 45.0 & 0.395 \\
50 & 160 & 41.2 & 0.421 \\
\noalign{\smallskip}\hline
\end{tabular}
\end{table}

The decrease in the CTT of Ag$_2$O may also be attributed to the enhanced lattice distortion of the film besides being due to the lowered activation energy of the reaction of film with hydrogen. Table \ref{tab:1} presents the average grain size and lattice strain of the Ag$_2$O film, as calculated in accordance with Williamson-Hall Relation \cite{20,21}:
\begin{equation}
\beta\cos\theta = \lambda /D + 4\times \Delta d/d \times \sin\theta
\label{eq:5}
\end{equation}
\begin{equation}
\beta^2 = {\beta_M}^2 - {\beta_S}^2,
\label{eq:6}
\end{equation}
where $\lambda$, $\theta$, $\it D$, $\it d$ and $\it {\Delta d/d }$ are the X-ray wavelength, diffraction angle, average grain size, interplanar spacing and lattice strain, respectively, $\beta$ is the full width at half maximum (FWHM) caused by the grain size and the lattice strain, and $\beta_M$ and $\beta_S$ are the measured FWHM of film and standard FWHM of the guide sample, respectively. The $\it D$ and $\it {\Delta d/d }$ can be obtained by the linear fitting of $\beta\cos\theta$ vs $\sin\theta$ for all the diffraction peaks in accordance with Williamson-Hall Relation, which can be automatically conducted by the software ( X'Pert HighScore Plus) accompanied with the XRD instrument (Philips PANAlytical X'Pert).The results indicate that a smaller average grain size leads to a larger lattice strain because of the enhanced grain surface effect. The lattice distortion is characterized by the gradually increasing lattice strain with increasing T$_a$ at different HPPs, resulting in a less compact film surface. In this case, the H$_2$ molecules adsorbed on the Ag$_2$O film surface may easily diffuse into the loose film and lead to an accelerated transformation of Ag$_2$O. To confirm this conclusion, the surface morphology of the Ag$_2$O film hydrogen annealed at different T$_a$ values using an HPP of 10\% was measured using a cold-field scanning electron microscope (Fig.\ref{fig:6e}).To obtain clear SEM images, carbon was sprayed on the surface of Ag$_2$O film to enhance the surface conductivity. The Ag$_2$O film hydrogen annealed
\begin{figure}
  \includegraphics[width=0.45\textwidth]{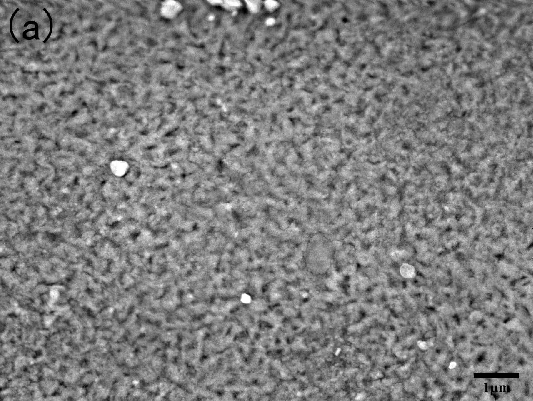}
\label{fig:6a}
%\caption{}
       % Give a unique label
\end{figure}
\begin{figure}
  \includegraphics[width=0.45\textwidth]{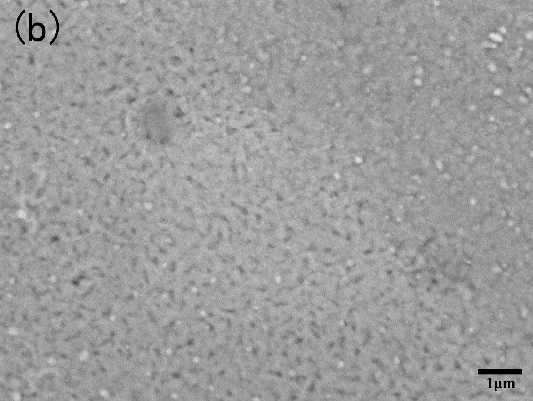}
\label{fig:6b}
%\caption{}
       % Give a unique label
\end{figure}
\begin{figure}
  \includegraphics[width=0.45\textwidth]{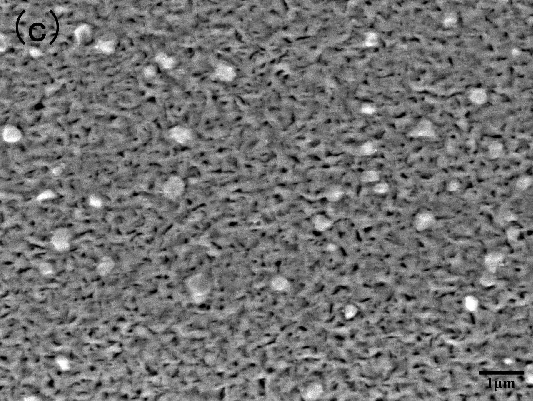}
\label{fig:6c}
%\caption{}
       % Give a unique label
\end{figure}
\begin{figure}
  \includegraphics[width=0.45\textwidth]{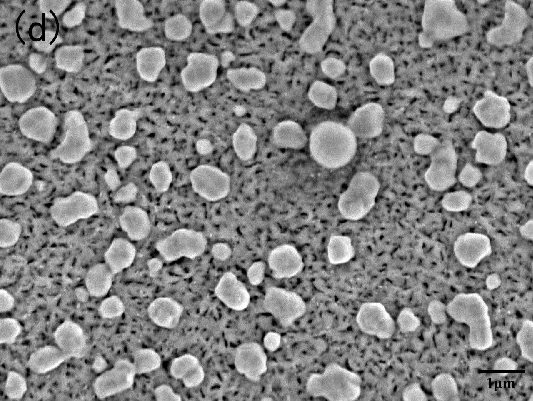}
\label{fig:6d}
%\caption{}
       % Give a unique label
\end{figure}
\begin{figure}
  \includegraphics[width=0.45\textwidth]{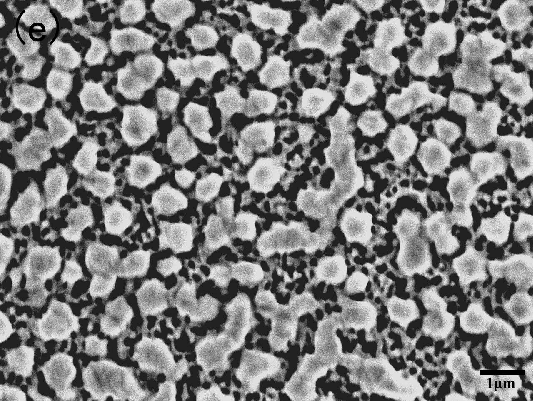}
\caption{Surface morphology of the Ag$_2$O films hydrogen annealed at temperature of (a) as-depo, (b) 160 $^o$C, (c) 175 $^o$C, (d) 200 $^o$C and (e) 225 $^o$C.}
       % Give a unique label
 \label{fig:6e}
\end{figure}
 at T$_a$ = 150 $^o$C shows a compact and uniform surface similar to the surface structure of the as-deposited Ag$_2$O film. However, the Ag$_2$O film hydrogen-annealed at T$_a$ = 175 $^o$C demonstrates a loose structure, and some holes begin to appear. According to Fig. \ref{fig:1}, Ag$_2$O begins to transform into Ag at T$_a$ = 175 $^o$C at an HPP of 10\%. The H$_2$ molecules adsorbed on the Ag$_2$O film surface may diffuse through the holes and accelerate the transformation of Ag$_2$O to Ag via the hydrogen reduction effect. Gaseous H$_2$O molecules may also diffuse to the film surface through these holes. After hydrogen annealing at 200 and 225 $^o$C, the Ag$_2$O film surface becomes much looser and more porous, again resulting in the accelerated transformation of Ag$_2$O to Ag via the hydrogen reduction effect. More gaseous H$_2$O molecules arising from the hydrogen reduction of the Ag$_2$O film may also diffuse to the film surface through these holes. Hence, increasing the T$_a$ and HPP may accelerate the transformation of the Ag$_2$O film. Increasing the HPP up to 30\% may lower the CTT of the Ag$_2$O film inherently because of the lowered activation energy of the reaction of Ag$_2$O with hydrogen and the enhanced lattice strain of Ag$_2$O film. It is worthy to be noted that in Figs.~\ref{fig:6e}a and ~\ref{fig:6e}c-e some agglomerates are discernible whereas none of them appears in Fig. 6b. The surface particle agglomerates may be due to a combination of different effects including not only the transformation of the film but also distortion of the SEM image at the pores. The inherent mechanism is still to be further studied.
\section{Conclusions}
Ag$_2$O films preferentially $<$111$>$ oriented  were deposited on glass substrates via DC magnetron sputtering and then annealed at different temperatures using different HPPs. The main results are as follows:
(1)	Increasing the HPP and T$_a$ can accelerate the transformation of the Ag$_2$O film because of the enhanced hydrogen reduction and the enhanced lattice distortion;
(2) Increasing the HPP (HPP $\le$ 30\%) can lower the CTT of the Ag$_2$O, which is inherently attributed to both the lowered activation energy of the reaction of Ag$_2$O with hydrogen and the enhanced lattice strain of Ag$_2$O film. When the HPP surpass 30\%, the CTT will remain unchanged even when HPP further increases.
(3) Hydrogen mainly acts as the reducing agent in the hydrogen annealing of the Ag$_2$O film. Hydrogen-involved reaction in the whole hydrogen annealing process is mainly hydrogen reduction reaction with Ag$_2$O. In addition, the diffusion of hydrogen and gaseous H$_2$O molecules is accompanied with the whole hydrogen annealing process.

%\begin{figure}
 %\includegraphics{fig_6a.eps}
%\end{figure}
%\begin{figure*}
 % \includegraphics[width=0.75\textwidth]{example.eps}
%\end{figure*}
%

\begin{acknowledgements}
I' m grateful for the supports from the National Natural Science Foundation of China (grant No. 60807001), the Foundation of Young Key Teachers from Univerisity of Henan Province (Grant No. 2011GGJS-008), the Foundation of Graduate Education Support of Zhengzhou University,the Foundation of Graduate Innovation of Zhengzhou University (Grant No. 12L00104) and the Foundation of Henan Educational Committee (Grant No. 2010A140017).
\end{acknowledgements}



\end{document}